\documentclass[pre,twocolumn, superscriptaddress,showpacs]{revtex4-1}%
\usepackage{amsfonts}
\usepackage{amsmath}
\usepackage{amssymb}
\usepackage{graphicx}%
\usepackage[english]{babel}
\usepackage{nicefrac}
\usepackage{ulem}

\usepackage[usenames,dvipsnames]{color}
\makeatletter
\@ifundefined{textcolor}{}
{%
 \definecolor{BLACK}{gray}{0}
 \definecolor{WHITE}{gray}{1}
 \definecolor{RED}{rgb}{1,0,0}
 \definecolor{GREEN}{rgb}{0,1,0}
 \definecolor{BLUE}{rgb}{0,0,1}
 \definecolor{CYAN}{cmyk}{1,0,0,0}
 \definecolor{MAGENTA}{cmyk}{0,1,0,0}
 \definecolor{YELLOW}{cmyk}{0,0,1,0}
 }

\newcommand{\beq}{\begin{equation}}
\newcommand{\eeq}{\end{equation}}
\newcommand{\nn}{\nonumber}
\newcommand{\ket}[1]{|#1\rangle}
\newcommand{\bra}[1]{\langle #1|}

\usepackage{babel}

\begin{document}

\title{Energy transport between two pure-dephasing reservoirs}

\author{T. Werlang
}\email{thiago\_werlang@fisica.ufmt.br}

\author{D. Valente
}\email{daniel@fisica.ufmt.br}


\affiliation{
Instituto de F\'isica, Universidade Federal de Mato Grosso, Cuiab\'a MT, Brazil}

\begin{abstract}

A pure-dephasing reservoir acting on an individual quantum system induces loss of coherence without energy exchange.
When acting on composite quantum systems, dephasing reservoirs can lead to a radically different behavior.
Transport of energy between two pure-dephasing markovian reservoirs is predicted in this work. 
They are connected through a chain of coupled sites. 
The baths are kept in thermal equilibrium at distinct temperatures. 
Quantum coherence between sites is generated in the steady-state regime and results in the underlying mechanism sustaining the effect.
A quantum model for the reservoirs is a necessary condition for the existence of stationary energy transport. 
A microscopic derivation of the non-unitary system-bath interaction is employed, valid in the ultrastrong inter-site coupling regime. The model assumes that each site-reservoir coupling is local.

\end{abstract}
\maketitle

%
\section{Introduction}

%
%
Quantum transport of energy and charge has been the subject of increasingly intense research during the past few years \cite{may}.
Advances in the fabrication of nanoscale systems and characterization of the fast dynamics of
a single or a few quantum systems now open wide avenues for understanding how
the laws of physics developed in the macroscopic domain are modified in the microscopic scale \cite{review}.
From the foundational perspective, a link between quantum dynamics and thermodynamic processes can
be envisioned \cite{nature.erez,briegel} and built \cite{ultracoldatoms,MolecularJunctionHeat}.
Biological transport processes can also be studied in the light of quantum mechanics, from coherent electron tunneling to
photosynthesis \cite{fleming, nori,vedral,NJP.castro}.
From the perspective of applications, quantum transport is basic to quantum information processing.
For instance, electronics can be performed with single electrons in quantum dots \cite{QD1}. 
This would allow the generation of large spin entangled currents in a passive device \cite{QD2}.
Quantum electronic transport has been recently reported in atomic-scale junctions \cite{MolecularJunction}.
Along with the electrons, heat flows through the junction. Unidirectional control of heat at the single-quantum level \cite{spinchains} is desirable in order to provide isolation of strategical centers in a circuit.
Quantum communication through photonic transport is promising due to the low decoherence suffered by light.
In that scenario, a quantum optical diode \cite{dudu} and the copying of a single-photon quantum state by a quantum emitter \cite{DV} have been proposed. Photon-mediated interaction between distant artificial atoms are now experimentally accessible \cite{science.wallraff}. 

%
%

Coherence on quantum transport is intimately related to the environment unavoidably coupled to the system of interest \cite{breuer}. The so called decoherence, or dephasing, induced by the environment is usually responsible for the loss of the quantum features in the dynamics of a microscopic system. Environments that suppress coherence while keeping unaltered the populations of the quantum states are called pure-dephasing reservoirs. Counterintuitively, recent attempts have drawn attention to the possibility of exploiting pure-dephasing environments as resources. Solid-state single-photon sources and nano-lasers have been proposed \cite{PureDephasingAsResource}, for instance. Pure-dephasing reservoirs have also been shown to boost quantum transport of energy in quantum networks, by destroying dark-states which effectively trap the excitation \cite{QD1,NJP.Plenio,PRB.Platero,PRB.Jaksch}. 
Quantum master equations techniques are largely employed,
where unitary and non-unitary dynamics are treated separately \cite{breuer}. The so called local approach consists in approximating the term describing the non-unitary dynamics of the ensemble by the sum of the terms describing the individual non-unitary processes. Such approach has been applied to the study of quantum transport, e.g., in Refs.\cite{NJP.Plenio,PRB.Jaksch,YingLi,PRB.Groth}. However, the local approach can lead to unphysical predictions, such as the violation of the second law of thermodynamics \cite{EPL.Korloff2ndLaw}.
Neglecting global non-unitary terms may also hide novel effects on the decay and dephasing rates of ultrastrongly-coupled systems \cite{PRA.Blais} and interference between independent reservoirs \cite{PRA.Lukin}.

%
%
In this paper, we microscopically model the effect of two independent pure-dephasing reservoirs 
individually coupled to each site of a two-site network and derive the global non-unitary dynamics of the network.
The main message we convey is that reservoirs which induce pure dephasing in the case of uncoupled sites can induce stationary energy exchange for coupled sites, due to the onset of quantum coherence in the steady-state. 
The physics reported herein deepen the knowledge on unexpected effects of dephasing reservoirs over the dynamics of quantum systems. In particular, Sec.\ref{HCSS} shows that energy current is given by a product of the inter-site coherence and the quantum nature of the bath, evidencing a genuinely quantum transport behavior.

The paper is organized as follows. In Sec. \ref{model}, we present the model and discuss possible physical implementations for it. Subsection \ref{effectiveenergyexchange} explores the dynamical origin of the energy exchange between system and bath. Sec. \ref{ME} shows the quantum master equation for the system dynamics, valid in the inter-site ultrastrong coupling regime.  Subsection \ref{effectivedecay} discusses a physical interpretation for the effective decay rates, making a link to the so-called Purcell effect. In Sec. \ref{HCSS}, the dynamics is calculated and the heat current is shown to be proportional to the quantum coherence between the sites. Sec.\ref{CLCA} evidences the limitations of the local approach and demonstrates that a classical reservoir does not provide stationary heat current through the chain.

%
%
\section{Model}
\label{model}

\begin{figure}
\begin{center}
\includegraphics[width=0.99\linewidth]{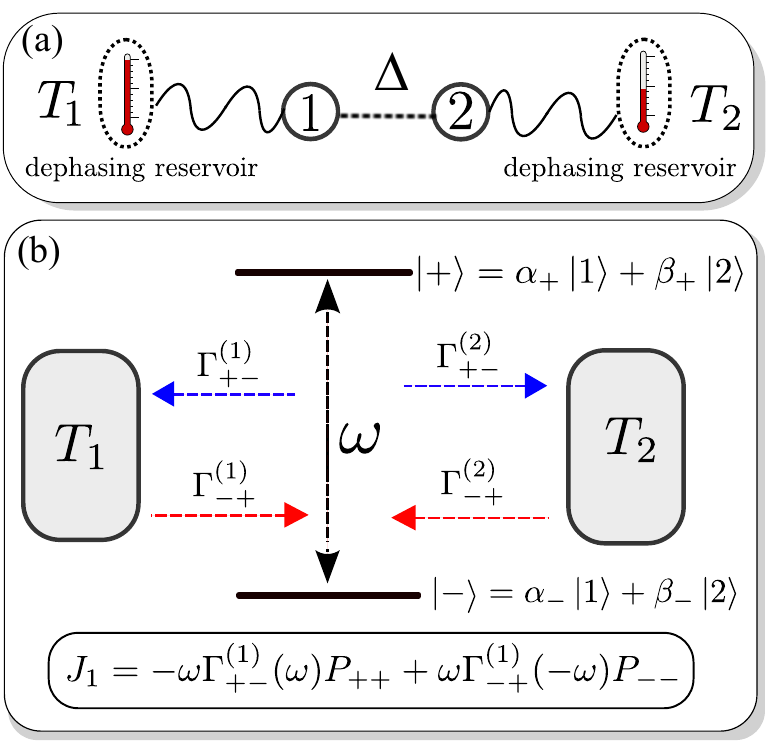}
\caption{
(a) Two sites coupled with strength $\Delta$. The site $1(2)$ is locally coupled to a thermal dephasing reservoir at temperature $T_{1(2)}$.
(b) $\ket{\pm}=\alpha_{\pm}\ket{1}+\beta_{\pm}\ket{2}$ are the eigenstates of $H_S$ with eigenvalues $\epsilon_{\pm}$. The energy transition is given by $\omega=\epsilon_+-\epsilon_-$.
$\Gamma^{(i)}_{+-}$ and $\Gamma^{(i)}_{-+}$ are the transition rates associated respectively with transitions $\ket{+}\rightarrow\ket{-}$ and $\ket{-}\rightarrow\ket{+}$, driven by the bath $i$. The energy current to/from bath $1$ is $J_1=-\omega\Gamma^{(1)}_{+-}(\omega)P_{++}+\omega\Gamma^{(1)}_{-+}(-\omega)P_{--}$, where $P_{\pm\pm}$ denotes the population of the eigenstate $\ket{\pm}$.}
\label{scheme}
\end{center}
\end{figure}

To highlight the quantum aspect of the energy transport we consider a system consisting of only two interacting sites, each of them coupled to a thermal dephasing reservoir. The system is illustrated in Figure \ref{scheme}(a). The two-site Hamiltonian is
\beq
H_S = h_1\ket{1}\bra{1} + h_2\ket{2}\bra{2} + \Delta(\ket{1}\bra{2}+\ket{2}\bra{1}),
\eeq
where $h_{1(2)}$ describes the energy of site $1(2)$ and $\Delta$ the coupling between the sites. State $\ket{i}$ denotes the presence
of an excitation at the {\it i}th site. As $\ket{1}\bra{1} + \ket{2}\bra{2}=\mathbf{1}$, the above Hamiltonian can be rewritten as follows
\beq\label{hsys}
H_S = h\ket{2}\bra{2} + \Delta(\ket{1}\bra{2}+\ket{2}\bra{1}),
\eeq
where $h=h_2-h_1$. The term proportional to the identity was disregarded, given that it is irrelevant for the system dynamics. The eigenvalues of $H_S$ are 
$\epsilon_{\pm}=\frac{1}{2}\left(h\pm\sqrt{h^2+4\Delta^2}\right)$ 
and the respective eigenstates are 
$\ket{\pm}=\alpha_{\pm}\ket{1}+\beta_{\pm}\ket{2}$, 
with 
$\alpha_\pm=\Delta/\sqrt{\Delta^2+\epsilon_\pm^2}$ 
and 
$\beta_\pm=\epsilon_\pm/\sqrt{\Delta^2+\epsilon_\pm^2}$.
Straightforward algebra shows that $\alpha_+ = -\beta_-$ and $\alpha_- = \beta_+$.
The motivation for choosing this model is the great variety of physical systems it describes. For Hubbard-type models of coupled quantum dots \cite{QD1,QD2}, $h$ is the on-site energy difference and $\Delta$ is the tunneling amplitude. For models of photosynthetic molecules \cite{nori}, $h$ is the energy difference between two chromophores and $\Delta$ is the excitonic coupling between them. The model can also describe the single-excitation subspace of a chain of coupled spins \cite{NJP.castro}.


The effects of the environment are taken into account by coupling each site to a bath of harmonic oscillators. 
The site-reservoir interaction is described by the Hamiltonian
\beq
H_{site-res}^{(i)} = \ket{i}\bra{i} \sum_k g_k (a_k^{(i)} + a_k^{(i)\dagger}), \quad i=1,2,
\label{Hint}
\eeq
with identical coupling strengths $g_k$. Note that it has the form of the so-called independent boson model \cite{mahan}, which describes, for instance, the interaction between a localized crystal defect and the lattice phonons field.
The Hamiltonians of the two free reservoirs are $H_{res,i} =  \sum_k \omega_k a_k^{(i)\dagger} a_k^{(i)}$. 
For noninteracting sites, $\Delta=0$, the dynamics induced by the reservoirs do not change the initial populations of the states $\ket{1}$ and $\ket{2}$, given that  $\left[H_S,H_{site-res}^{i}\right]=0$.
The bath induces only decoherence, at a rate 
$
\gamma_\phi=\lim_{\nu_0\rightarrow 0}\sum_{\nu,i} g^2_\nu\ n^{(i)}_\nu \delta(\nu-\nu_0)
=\lim_{\nu\rightarrow 0}\frac{\mathcal{J}(\nu)}{\nu} k_B\bar{T}
$, 
where $n^{(i)}_\nu$ is the Bose-Einstein distribution (see Eq.\ref{BE}), $\bar{T}$ is the average temperature of the reservoirs and $\mathcal{J}(\nu)$ is the spectral density of the bath (see section \ref{LDM} for details). 
In this sense, we have a typical pure-dephasing reservoir.
On the other hand, for a finite coupling $\Delta$, the baths modeled by Eq. (\ref{Hint}) induce not only decoherence, but also relaxation, as discussed below.

\subsection{Effective energy-site exchange}
\label{effectiveenergyexchange}

The counterintuitive effect of energy exchange between a site and a pure-dephasing bath becomes evident when the system operator coupled to the bath is written in the $H_S$ eigenstate basis, namely,
\beq
\ket{1}\bra{1} = \alpha_+^2 \ket{+}\bra{+}+\alpha_-^2 \ket{-}\bra{-}
+ \alpha_+\alpha_- (\ket{+}\bra{-}+\ket{-}\bra{+})
\nn
\eeq
and
\beq
\ket{2}\bra{2} = \alpha_-^2 \ket{+}\bra{+}+ \alpha_+^2 \ket{-}\bra{-}
- \alpha_+ \alpha_- (\ket{+}\bra{-}+\ket{-}\bra{+}).
\nn\eeq
In each equation, the first two operators describe authentic pure-dephasing, whereas the third term gives rise to energy exchange between system and bath, $H_{SB-Exch}$, with an effective coupling proportional to the product 
$\alpha_+ \alpha_-$, 
i.e.,
\beq
H_{SB-Exch}^{(i)} = \alpha_+ \alpha_- (\ket{+}\bra{-}+\ket{-}\bra{+}) \sum_k g_k (a_k^{(i)} + a_k^{(i)\dagger}).
\label{effEx}
\eeq
Note that in the usual weak-coupling limit, $\Delta \ll h$, the effective energy-exchange coupling vanishes linearly, $\alpha_+ \approx \Delta/h + \mathcal{O}(\Delta/h)^3$ and $\alpha_- \approx 1 + \mathcal{O}(\Delta/h)^2$, hence $\alpha_+ \alpha_- \approx \Delta/h$.

\section{Master equation in the ultrastrong-coupling formalism}
\label{ME}

To determine the dynamics of the two interacting sites we assume that the coupling between the sites and the reservoirs is weak. However, it is important to mention that there is no restriction with respect to the coupling between sites. Thus, our results can be used even in the ultrastrong coupling regime ($\Delta \gtrsim h$) \cite{ultrastrong}. The dynamics of the system of interest can be deduced from the complete system-plus-bath Hamiltonian, leading to the following quantum Markovian master equation \cite{breuer}
\beq
\frac{d\rho}{dt} =-i [H_S,\rho] + \mathcal{L}_1[\rho] + \mathcal{L}_2[\rho],
\label{master}
\eeq
in $\hslash = 1$ units, where the Lindblad superoperators $\mathcal{L}_i[\rho]$ ($i=1,2$) are given by
\begin{eqnarray}
{\cal L}_{i}[\rho] & = & \sum_{\nu} \gamma_{i}(\nu)
\bigg[ A_{i}(\nu)\rho A_{i}^{\dagger}(\nu) - \frac{1}{2}\Big\{ \rho,A_{i}^{\dagger}(\nu)  A_{i}(\nu) \Big\} \bigg] 
\label{Lindbladians}
\end{eqnarray}
where $\nu=\epsilon - \epsilon'$. Here $\epsilon$ and $\epsilon'$ are two arbitrary eigenvalues of $H_S$.
All the properties of the reservoir are contained in $\gamma_{i}(\nu)$. For a quantum heat bath of harmonic oscillators at a temperature $T_i$, we have that $\gamma_{i}(\nu)=\mathcal{J}_i(\nu) (1+n^{(i)}_\nu)$ for $\nu>0$ and $\gamma_{i}(\nu)=\mathcal{J}_i(\nu)n^{(i)}_{|\nu|}$ for $\nu<0$, where $\mathcal{J}_i(\nu)$ is the bath spectral density. The average number of excitations $n^{(i)}_\nu$ at temperature $T_i$ in the {\it i}th reservoir is given by the Bose-Einstein distribution
\beq
n^{(i)}_\nu = \left[\exp{\frac{\nu}{k_B T_{i}}}-1\right]^{-1}.
\label{BE}
\eeq
The Lindblad operator associated with the {\it i}th reservoir is
\[
A_{i}(\nu)=\sum_{\nu=\epsilon-\epsilon'}\Pi_{\epsilon'}\ket{i}\bra{i}\Pi_{\epsilon}
\]
where $\Pi_{\epsilon}$ the projection onto the eigenspace belonging to the eigenvalue $\epsilon$. The operator $A_{i}(0)$ describes the dephasing effects due to the interaction with the {\it i}th reservoir, while $A_{i}(\nu)$ is related to the transition between the eigenstates with energy gap equal to $\nu\neq0$. In our case, the sum is made over $\nu=0, \pm\omega$, where $\omega= \epsilon_+-\epsilon_-=\sqrt{h^2+4\Delta^2}$, and $\Pi_{\epsilon_\pm}=\ket{\pm}\bra{\pm}$. Therefore,      
\begin{eqnarray*}
A_{1}(0)&=&\alpha_-^2\ket{-}\bra{-}+\alpha_+^2\ket{+}\bra{+},\\
A_{2}(0)&=&\beta_-^2\ket{-}\bra{-}+\beta_+^2\ket{+}\bra{+},\\
A_{1}(\omega)&=&\alpha_+\alpha_-\ket{-}\bra{+},\\
A_{2}(\omega)&=&\beta_+\beta_-\ket{-}\bra{+},
\end{eqnarray*}
with $A_{i}^\dagger(\omega)=A_{i}(-\omega)$. As already stated, $\alpha_{\mp} = \pm \beta_{\pm}$. Using these results, Eq. (\ref{Lindbladians}) takes the form
\begin{eqnarray}
{\cal L}_{i}[\rho] & = & \gamma_{i}(0)
\bigg[ A_{i}(0)\rho A_{i}(0) - \frac{1}{2}\Big\{ \rho,A_{i}(0)  A_{i}(0) \Big\} \bigg]\nonumber\\
& + & \Gamma^{(i)}_{+-}
\bigg[ \ket{-}\bra{+}\rho \ket{+}\bra{-} - \frac{1}{2}\Big\{ \rho,\ket{+}\bra{+} \Big\} \bigg] \\
& + & \Gamma^{(i)}_{-+}
\bigg[ \ket{+}\bra{-}\rho \ket{-}\bra{+} - \frac{1}{2}\Big\{ \rho,\ket{-}\bra{-} \Big\} \bigg] \nonumber,
\label{Lindbladians1}
\end{eqnarray}
where 
$
\gamma_i(0)=\lim_{\nu\rightarrow0}\gamma_i(\nu)
=
\lim_{\nu\rightarrow0} \frac{\mathcal{J}(\nu)}{\nu} k_BT_i
$ 
is the dephasing rate due to the $i$th reservoir. 
The transition rates of transitions 
$\ket{+}\rightarrow\ket{-}$ 
and
$\ket{-}\rightarrow\ket{+}$ 
are
$\Gamma^{(i)}_{+-} (\omega)=\gamma_{i}(\omega)(\alpha_+  \alpha_-)^2$
and
$\Gamma^{(i)}_{-+} (\omega)=\gamma_{i}(-\omega)(\alpha_+  \alpha_-)^2$,
respectively, where $\gamma_{i}(\pm \omega)$ has been written above.

\subsection{Effective decay rate in the ultrastrong-coupling formalism}
\label{effectivedecay}

The effective decay rate 
$\Gamma^{(i)}_{+-} (\omega)$, 
derived microscopically from $H_{site-res}$ in Eq.(\ref{Hint}), deserves further analysis.
The original expression for the decay rate,
\beq
\Gamma^{(i)}_{+-}(\omega) = 
\sum_k (\alpha_+ \alpha_-)^2\ g_k^2 \ (n_{\nu_k}^{(i)}+1)\  \delta(\nu_k - \omega),
\label{gammaeff}
\eeq 
evidences the influence of the coupling $\Delta$ between the $i$-th site and its neighbor, 
as written in $(\alpha_+\alpha_-)^2$,
on the damping that emerges from the coupling $g_k$ between the $i$-th reservoir and the $i$-th site itself.
Note that it is a consequence of the effective site-bath energy exchange Hamiltonian, $H_{SB-Exch}$, in Eq.(\ref{effEx}).
Here, we call attention to the fact that dissipative dynamics in open quantum systems are essentially encoded in the bath spectral density $\mathcal{J}_i(\nu)$, with which the continuum limit is computed, $\sum_k \rightarrow \int d\nu \mathcal{J}_i (\nu)$.
Eq.(\ref{gammaeff}) suggests that the neighboring site is effectively altering the spectral density from $i$-th bath, 
$\mathcal{J}_i(\nu)\rightarrow \tilde{\mathcal{J}}_i(\nu)$. 
We define the effective spectral density such that
$\Gamma^{(i)}_{+-}(\omega) =
(\alpha_+ \alpha_-)^2 \mathcal{J}_i(\omega) (n_{\omega}^{(i)}+1)
\equiv 
\tilde{\mathcal{J}}_i(\omega)  (n_{\omega}^{(i)}+1)$,
where 
$\omega=\epsilon_+ - \epsilon_-$ 
is the gap. 
The explicit dependence of $(\alpha_+ \alpha_-)^2$ on the gap is found, 
$\alpha_+ \alpha_- = \Delta/\omega$, 
hence the effective spectral function reads
\beq
\tilde{\mathcal{J}}_i(\omega) = \mathcal{J}_i(\omega)\  \frac{\Delta^2}{\omega^2}.
\label{effJ}\eeq
The coupling to the neighboring site introduces, thus, a sub-ohmic correction on the free bath spectral function.
Modification of the decay rate due to an alteration of the spectral function of the reservoir
happens in the well-known Purcell effect \cite{purcell}. For instance, spontaneous emission
of an atom in free space can be accelerated if the emitter is put between two mirrors \cite{purcellHarocheJMG}.
In that case, the alteration of the spectral function comes from the structured environment itself, as 
the mirrors change the density of electromagnetic modes available to the emitter.
Eq.(\ref{effJ}) reveals analogous effect, of a rather different origin, though.
Here, the decay of energy from one site to the reservoir with which it is coupled is being affected by 
the coupling of such site to a neighboring one.
This result suggests that the emission rate of an atom coupled to a heat bath can be modified 
due to the coupling not only with a cavity, but also with another atom. 
By an abuse of terminology, this could be said to be a kind of fermionic Purcell effect, 
in contrast to the standard bosonic one.
This is a possible research perspective opened by the present study.

%
%
\section{Heat current in the steady-state regime}
\label{HCSS}

The central point to derive an expression for the heat current $J_\mathrm{heat}$ in a quantum system is to relate the average energy going through the system $\langle H_S\rangle$ with the continuity equation \cite{briegel},
\beq
\frac{\partial}{\partial t} \langle H_S \rangle = -\nabla \cdot J_\mathrm{heat} = - (J_2 - J_1).
\label{cont}
\eeq
Using Eq. (\ref{master}) and noting that
$
\frac{\partial}{\partial t} \langle H_S \rangle = \frac{\partial}{\partial t} \mbox{Tr}\{\rho  H_S\} = \mbox{Tr}\{\dot{\rho}H_S\},
$
the left-hand side of Eq. (\ref{cont}) can be rewritten as
\beq
\frac{\partial}{\partial t} \langle H_S \rangle = \mbox{Tr}\{\mathcal{L}_L[\rho] H_S\}  + \mbox{Tr}\{\mathcal{L}_R[\rho] H_S\}. 
\label{heatinout}
\eeq
Comparing the right-hand sides of Eqs. (\ref{cont}) and (\ref{heatinout}) we can write the input energy rate $J_{1(2)}^{\mathrm{in}}$ from the reservoir $1(2)$ as
\beq
J^\mathrm{in}_{1(2)} \equiv \mbox{Tr}\{\mathcal{L}_{1(2)}[\rho] H_S\} = \pm J_{1(2)}.
\label{jin}
\eeq
For the system under study, this gives 
$
J_1=-\Gamma^{(1)}_{+-}(\omega)(\epsilon_+-\epsilon_-)P_{++}
+\Gamma^{(1)}_{-+}(-\omega)(\epsilon_+-\epsilon_-)P_{--},
$
where $P_{++}$ and $P_{--}$ are respectively the excited and ground state populations, as computed in the following.
It is worth pointing out that heat current is defined here in agreement to the 1st Law of Thermodynamics, in its generalized version to open quantum systems \cite{PRE.Esposito,arxiv.thermo}. In Ref.\cite{arxiv.thermo}, a phenomenological modeling of the reservoirs precludes one from defining temperature and, therefore, heat flow. In contrast, our microscopic modeling of thermal equilibrium reservoirs allows us to treat energy flow as heat flow. Besides, it assigns physical meaning to the derived rates $\gamma_i(\nu)$.

Heat current in the steady-state regime is obtained by Eq. (\ref{jin}), using the stationary solution of Eq. (\ref{master}), that is, $\dot{\rho}_{ss}=0$. In this case, $J_1=-J_2$ because $\frac{\partial}{\partial t} \langle H_S \rangle = 0$. The time-dependent solution of Eq. (\ref{master}) in the eigenstate basis $\ket{\pm}$ is given by
\begin{eqnarray}
P_{++}(t)=&\frac{\Gamma^{tot}_{-+}}{\Gamma^{tot}_{+-}+\Gamma^{tot}_{-+}}&\nn\\
&+\left(\frac{\Gamma^{tot}_{+-}}{\Gamma^{tot}_{+-}+\Gamma^{tot}_{-+}}-P_{--}(0)\right)e^{-(\Gamma^{tot}_{+-}+\Gamma^{tot}_{-+})t}, \nonumber\\
P_{--}(t) =& \frac{\Gamma^{tot}_{+-}}{\Gamma^{tot}_{+-}+\Gamma^{tot}_{-+}}& \nn\\ 
&-\left(\frac{\Gamma^{tot}_{+-}}{\Gamma^{tot}_{+-}+\Gamma^{tot}_{-+}}-P_{--}(0)\right)e^{-(\Gamma^{tot}_{+-}+\Gamma^{tot}_{-+})t},\nn\\
P_{+-}(t)=&P_{+-}(0)e^{-i\Delta t}e^{-\frac{1}{2}(\Gamma^{tot}_{+-}+\Gamma^{tot}_{-+}+\gamma_\phi)t}&,\nonumber
\end{eqnarray}
where $P_{\pm\pm}=\bra{\pm}\rho\ket{\pm}$ and
\begin{eqnarray}
\Gamma^{tot}_{+-}&=& \Gamma^{(1)}_{+-}+\Gamma^{(2)}_{+-}\nonumber\\
\Gamma^{tot}_{-+}&=& \Gamma^{(1)}_{-+}+\Gamma^{(2)}_{-+}\\
\gamma_\phi&=&\gamma_1(0)(\alpha_+^2-\alpha_-^2)^2 +\gamma_2(0)(\beta_+^2-\beta_-^2)^2.\nonumber
\end{eqnarray}
Therefore, in the steady-state regime, $ (\Gamma^{tot}_{+-}+\Gamma^{tot}_{+-})\ t\gg 1$, the density matrix is
\beq 
\rho_{ss}=\frac{\Gamma^{tot}_{-+}}{\Gamma^{tot}_{+-}+\Gamma^{tot}_{-+}}\ket{+}\bra{+} + \frac{
\Gamma^{tot}_{+-}}{\Gamma^{tot}_{+-}+\Gamma^{tot}_{-+}}\ket{-}\bra{-},
\label{rhoss1}\eeq
that can be computed in terms of $\bar{n}=(n^{(1)}_\omega+n^{(2)}_\omega)/2$, 
and 
$\delta n=n^{(2)}_\omega-n^{(1)}_\omega$, 
as
\beq
\rho_{ss}=\frac{\bar{n}}{2\bar{n}+1}\ket{+}\bra{+} 
+ \frac{\bar{n}+1}{2\bar{n}+1}\ket{-}\bra{-}.
\label{rhoss}
\eeq

Applying the steady-state found above, the current becomes
\beq
J_1=
-\mathcal{\tilde{J}}_1(\omega) 
\left(\epsilon_+-\epsilon_-\right)\left(P_{--}-P_{++}\right)\frac{\delta n}{2},
\label{heatcurrent}\eeq
which evidences the linear dependence on the gradient of the baths average excitations, $\delta n$.
By noting that
\beq
P_{--}-P_{++} = -\frac{\rho_{12}}{\alpha_+ \alpha_-},
\eeq
the genuine quantum features of $J_1$ become clear. 
Firstly, the quantum signature of the system is in $\rho_{12}$, i.e., the quantum coherence between sites $1$ and $2$.
Remarkably, such quantum coherence is generated by the pure-dephasing reservoirs, and maintained in the steady-state regime. A quantum signature of the bath is on the spontaneous emission of energy from the system to the bath, with rate  $\tilde{\mathcal{J}}_1(\omega)$. Additionally, the nonlinearity of $\delta n$ with respect to the temperature gradient indicates the threshold between quantum and classical regimes for the reservoirs, as discussed below in Eq.(\ref{nonlinn}). Heat current is proportional to the product of these three quantities. Therefore, both system and baths must be in a quantum regime in order to trigger energy transport between reservoirs.

\begin{figure}
\begin{center}
\includegraphics[width=0.99\linewidth]{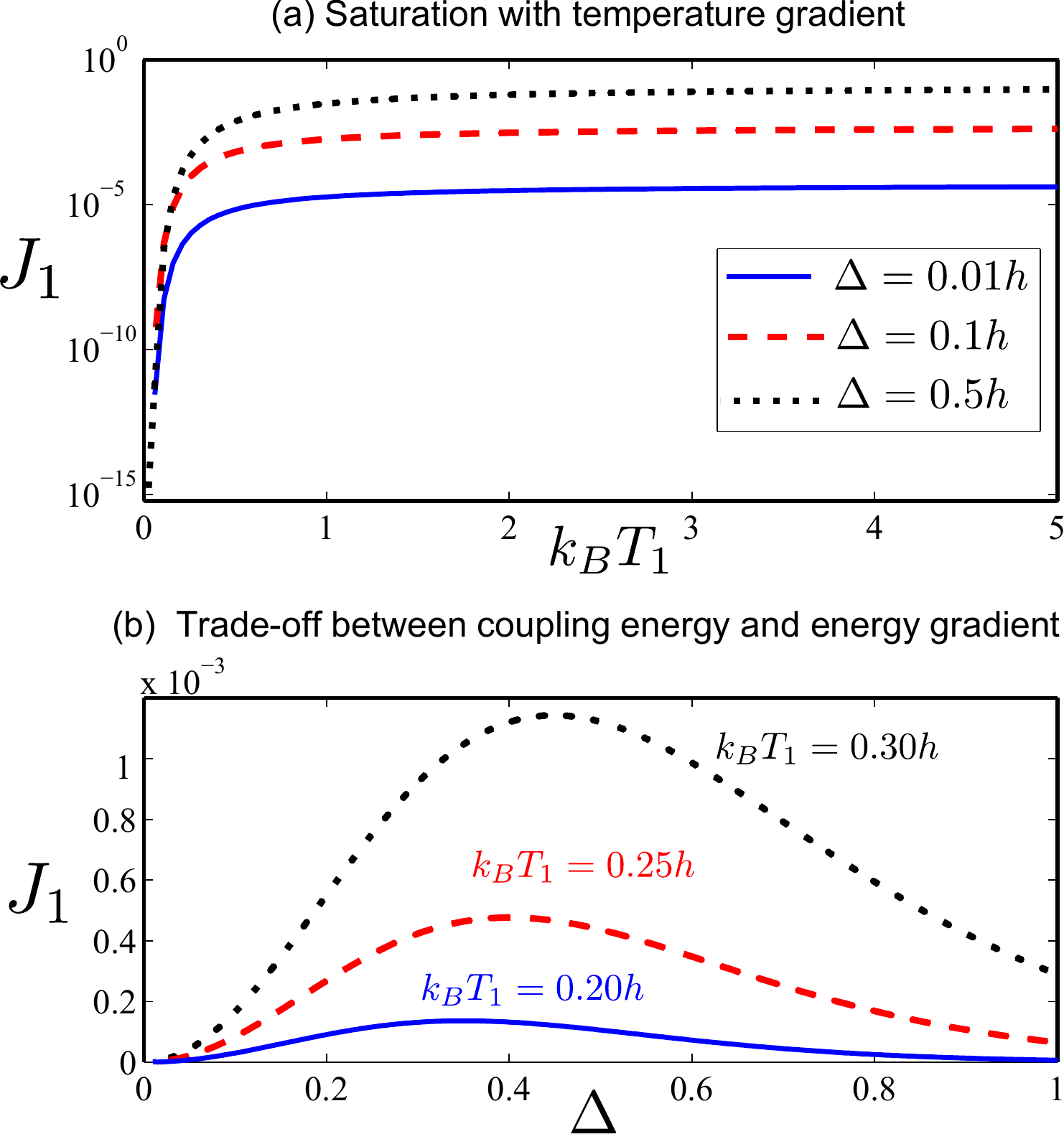}
\caption{
(a) The heat current $J_1$ in the steady-state regime as a function of the temperature $k_BT_1$ for $k_BT_2=0.01h$ and $\Delta=0.01h$ (blue solid line), $\Delta=0.1h$ (red dashed line), and $\Delta=0.5h$ (black dotted line). The heat current saturates at the value $J_1= \kappa \Delta^2 /4$ as the temperature $k_BT_1$ increases. We adopt an ohmic spectral density $\mathcal{J}(\omega)=\kappa \omega$ with $\kappa=1$. (b) The heat current $J_1$ as a function of the energy coupling $\Delta$ for $k_BT_2=0.1h$ and $k_BT_1=0.2h$ (blue solid line), $k_BT_1=0.25h$ (red dashed line), and $k_BT_1=0.3h$ (black dotted line). The heat current $J_1$ tends to zero in the low temperature limit, that is, $\omega=\sqrt{h^2+4\Delta^2}\ggg k_BT_1,k_BT_2$. For a fixed temperature gradient, this limit can always be achieved by increasing the energy coupling $\Delta$.}
\label{fig2}
\end{center}
\end{figure}

The behavior of the heat current $J_1$ is illustrated in Fig.\ref{fig2}. Saturation of the current with temperature gradient,  Fig.\ref{fig2}(a), occurs as a consequence
of the product between an increasing temperature gradient, which makes $\delta n$ to increase, and a
decreasing coherence between sites, $\rho_{12}$. In the limit $n_\omega^{(1)}\gg 1 \gg n_\omega^{(2)}$,
the saturating current is given by $J_1 \rightarrow \mathcal{\tilde{J}}(\omega)\ \omega /4$. The choice of an
ohmic spectral density, $\mathcal{J}(\omega)=\kappa \omega$, implies that the saturation current 
depends only on the coupling $\Delta$, not on the gap $\omega$, $J_1 \rightarrow \kappa \Delta^2 /4$. On the other hand,  Fig.\ref{fig2}(b) shows that for a fixed temperature gradient, the heat current $J_1$ first grows to a maximum value and then decreases as the energy coupling $\Delta$ increases. To understand this behavior, note that for sufficiently large values of $\Delta$, that is,  in the low temperature limit $\omega=\sqrt{h^2+4\Delta^2}\ggg k_BT_1,k_BT_2$, we have that $\delta n\approx0$ and $\bar{n}\approx0$. Consequently, since in this limit the steady state is close to $\ket{-}$, the heat current becomes $J_1\rightarrow-\kappa\Delta^2\delta n/2\approx0$. 

In the classical Fourier law for heat conduction \cite{ClassFourierLaw}, 
heat current is linearly proportional to the temperature gradient, 
$J_{Fourier} \propto - \nabla T = -(T_2-T_1)$.
Heat current in Eq.(\ref{heatcurrent}) can be regarded as a generalization of that law,
in the sense that the current is propotional to the number gradient, 
$J_1 \propto -\delta n = -(n^{(2)}_\omega-n^{(1)}_\omega)$. 
For high temperatures $k_B T_{1,2}\gg \omega$, though,
\begin{eqnarray}
\delta n \equiv & n^{(2)}_\omega-n^{(1)}_\omega & \nn\\
=& \frac{1}{e^{\frac{\omega}{k_B T_2}}-1}-\frac{1}{e^{\frac{\omega}{k_B T_2}}-1}\nn \\
\approx &\frac{k_B T_2}{\omega}-\frac{k_B T_1}{\omega} \nn\\
=& \frac{k_B}{\omega} (T_2-T_1),
\label{nonlinn}
\end{eqnarray}
showing that such generalization recovers the linearity on the temperature gradient
in the high temperature (classical) limit. 
Withing the same approximations, it is found that the steady-state, Eq.(\ref{rhoss}), depends only on the average temperature, 
$\bar{n}\equiv(n^{(2)}_\omega + n^{(1)}_\omega)/2\approx \frac{k_B}{\omega}(\frac{T_1+T_2}{2})$,
not on the temperature gradient.

To conclude this section, we remark the difference between the effect presented in this paper and dephasing-induced energy transport effects already reported elsewhere \cite{NJP.Plenio,PRB.Jaksch,YingLi}.
The usual scenario is the following. The establishment of quantum coherence imposes some kind of insulation, locking the excitation in the chain. The role of noise is to unlock excitation flow by breaking quantum coherence and, thus, to induce the suppression of the inefficient pathways.
Our model shows two dissimilar properties: firstly, the pure-dephasing reservoir builds quantum coherence instead of suppressing it, and secondly, quantum coherence creates energy flow, instead of locking it.

%
%
\section{Comparison to local and classical approaches}
\label{CLCA}

In the following, we show that (A) the predictions of the local modeling are unphysical in the low-temperature regime and
(B) that heat current in the steady-state is established only if thermal baths are modeled in a quantum mechanical framework.

\subsection{Local dephasing model}
\label{LDM}

In the case of two weakly interacting sites ($\Delta\lll h$), it is common to use a local approach for the energy transport in quantum systems. Such an approach ignores the effects of coupling between the sites in the description of dissipative dynamics. Thus, the system dynamics is governed by the master equation (\ref{master}) with the phenomenological Lindblad superoperators 
\begin{eqnarray}
{\cal L}^{ph}_{i}[\rho] & = & \gamma_{i}(0)\bigg[ \ket{i}\bra{i}\rho\ket{i}\bra{i} - \frac{1}{2}\Big\{ \rho,\ket{i}\bra{i} \Big\} \bigg].
\label{Lindph}
\end{eqnarray}
In the site basis $\left\{\ket{1},\ket{2}\right\}$,
\begin{eqnarray}
\dot{\rho}_{11} &=& -i\Delta\left(\rho_{21}(t)-\rho_{12}(t)\right),\nonumber\\
\dot{\rho}_{22} &=& i\Delta\left(\rho_{21}(t)-\rho_{12}(t)\right),\\
\dot{\rho}_{12} &=& -i\Delta\left(\rho_{22}(t)-\rho_{11}(t)\right)
+ih\rho_{12}(t)+\frac{1}{2}\gamma_\phi^{ph}\rho_{12}(t),\nonumber,
\end{eqnarray}
where $\gamma_\phi^{ph}=\gamma_1(0)+\gamma_2(0)$. 

The heat current in the local approach reads
\beq
J_1^{ph} = -\gamma_1(0)\  \Delta\ \mbox{Re}[\rho_{12}],
\label{Jphen}
\eeq
also depending crucially on coherence between sites.

The steady-state within the local approach is
\beq
\rho_{ss}^{ph}=\frac{1}{2}\left(\ket{1}\bra{1}
+\ket{2}\bra{2}\right)=\frac{1}{2}\left(\ket{-}\bra{-}+\ket{+}\bra{+}\right),
\label{ssph}\eeq 
for which $\rho_{12} = 0$, so $J_1^{ph}=0$, as well. That is, the local approach does not capture
steady-state flow of heat between pure-dephasing reservoirs.


Note, however, that the steady-state resulting from the local approach contains 
unphysical predictions in the low temperature regime. 
Take, for instance, $k_B T, \Delta \lll h$, with $\rho(0)=\ket{1}\bra{1}$ as the initial state.
Because $\Delta \lll h$, the ground state of the system is arbitrarily close to $\ket{1}$.
The arbitrarily low temperature $k_B T \lll h$ guarantees that thermal jumps from the ground to the excited
state, $\ket{1}\rightarrow\ket{2}$, 
occur with vanishing probability, $p_{1\rightarrow2}/p_{2\rightarrow 1}\sim \exp-h/k_B T \lll 1$.
Vanishing temperatures also imply vanishingly small dephasing rate, $\gamma_\phi^{ph}\propto k_B T \lll h$,
hence the dynamics is arbitrarily close to unitary.
Were the dynamics unitary, the population of state $\ket{1}$ would evolve as 
$\rho_{11}(t)=1-2\frac{\Delta^2}{\omega^2}(1-\cos(\omega t))$, 
which deviates from $\rho_{11}(t)\approx 1$ by a factor of $\sim 4\Delta^2/h^2 \lll 1$.
Therefore, neither unitary nor non-unitary dynamics are expected to void the system from state 
$\ket{1}\bra{1}$ with finite probability.
In clear contrast to the intuitively expected state, 
the locally derived steady-state of Eq.(\ref{ssph}) is a mixture of $\ket{1}\bra{1}$ and $\ket{2}\bra{2}$ with
precisely the same weights.

The microscopic derivation does not suffer from this pathology.
For two reservoirs at the same temperature $T$ (i.e., $\delta n = 0$), 
the steady-state consists in thermal equilibrium, or the Gibbs state, of the global system, $\rho_{ss}\rightarrow \rho_{T}$, 
\beq
\rho_{T} = \frac{e^{-\beta H}}{\mbox{Tr}[e^{-\beta H}]} 
= \frac{n}{2{n}+1}\ket{+}\bra{+} + \frac{{n}+1}{2{n}+1}\ket{-}\bra{-} = \rho_{ss},
\label{gibbs}
\eeq
where Eq.(\ref{rhoss}) has been applied, along with $n = [\exp{(\beta \omega)}-1]^{-1} = \bar{n}$, 
and $\beta=1/(k_B T)$. In the vanishing temperature limit, it then simplifies to $\rho_T \rightarrow \ket{-}\bra{-}$.
$\Delta \lll h$ implies that $\ket{-}\rightarrow \ket{1}$, so 
\beq
\rho_{T\rightarrow 0}\big|_{\Delta \lll h} \approx \ket{1}\bra{1},
\nn\eeq
as intuitively expected.

It is important to underline that $\rho_{ss}^{ph}$ coincides with $\rho_{T}$ 
in the high temperature limit, $k_B T \gg \omega = \sqrt{h^2+4\Delta^2}$, 
for which the Gibbs state is a complete mixture of the ground and the excited state, 
$\rho_{T\rightarrow\infty} \approx \frac{1}{2}\ket{+}\bra{+}+\frac{1}{2}\ket{-}\bra{-}$, for arbitrary $\Delta$ and $h$.


Energy flow between a two-level system and a pure-dephasing bath led by coherence has been recently reported in Ref \cite{arxiv.thermo}. However, in that case the local approach has been applied without any microscopic derivation and a heat current similar to Eq.(\ref{Jphen}) has been derived. As it has just been shown above, such modeling predicts energy flow in the transient regime, but not in steady-state, for which coherence vanishes.

Using Many-Body Green's functions techniques, the authors of Ref.\cite{EPL.Wu} have recently studied heat flow in a spin-boson nanojunction. Their approach is valid for arbitrary system parameters and spin-bath couplings. Whereas in their model the two reservoirs are coupled to a single spin, we study a local site-bath coupling. It is also worth emphasizing that our Quantum Master Equation approach is particularly useful to identify the equilibration dynamics of the two-site chain towards thermal Gibbs state, as shown by Eqs.(\ref{rhoss}) and (\ref{gibbs}). Moreover, it highlights the effective decay rate, due to the inter-site coupling, that provides the timescale for attaining equilibrium. 

\subsection{Classical dephasing model}

Eq. (\ref{heatcurrent}) indicates that the quantum nature of the reservoir is crucial to the emergence of the heat current between pure-dephasing baths. To investigate this point more carefully, each site is now coupled to a classical dephasing reservoir. For this purpose, instead of a set of harmonic oscillators, the reservoir is modeled by a stochastic function of time, which describes general energy fluctuations \cite{PRA.Blais}. The site-reservoir Hamiltonian is
\beq
H_{site-res}^{(i)} = \ket{i}\bra{i} f_i(t), \quad i=1,2,
\label{Hintc}
\eeq
where $f_i(t)=\int_{-\infty}^\infty \tilde{f}_i(\nu)e^{i\nu t}d\nu$ is a stochastic function of time, with $\left\langle \tilde{f}_i(\nu)\right\rangle_c=0$ and $\left\langle \tilde{f}_i(\nu)\tilde{f}_i(-\nu')\right\rangle_c=S_i(\nu)\delta(\nu-\nu')$ \cite{PRA.Blais}. Here $\left\langle \cdot\right\rangle_c$ denotes the classical average and $S_i(\nu)$ the spectral density of $f_i(t)$.

The same microscopic approach used in the quantum case can be applied to derive a Markovian master equation in the classical case \cite{PRA.Blais}. The dynamics of the system is obtained again using Eqs. (\ref{master})-(\ref{Lindbladians}), with $\gamma_i^c(\nu)$ instead of $\gamma_i(\nu)$. The difference between the quantum and the classical baths appears only in the function $\gamma_i^c(\nu)$, which describes the characteristics of a classical dephasing reservoir. In this case,
\begin{eqnarray*}
\gamma_i^c(\nu)&\equiv&\int_{-\infty}^\infty d\tau e^{i\nu \tau} \left\langle f_i(\tau)f_i(0)\right\rangle_c \\
&=& \int_{-\infty}^\infty d\tau\int_{-\infty}^\infty d\nu'\int_{-\infty}^\infty d\nu'' e^{i(\nu+\nu') \tau}\left\langle \tilde{f}_i(\nu')\tilde{f}_i(\nu'')\right\rangle_c\\
&=&2\pi\int_{-\infty}^\infty d\nu''\left\langle \tilde{f}_i(-\nu)\tilde{f}_i(\nu'')\right\rangle_c\\
&=&2\pi S_i(-\nu).
\end{eqnarray*}
Furthermore, as $S_i(\nu)=S_i(-\nu)$ \cite{gardiner}, we have that $\gamma_i^c(-\nu)=\gamma_i^c(\nu)$. This result shows that the classical version of the transition rates $\Gamma_{+-}^{(i)}(\omega)$ and $\Gamma_{-+}^{(i)}(\omega)$ are equal, because $\Gamma_{+-}^{(i)}(\omega)-\Gamma_{+-}^{(i)}(-\omega)=\left(\alpha_+\alpha_-\right)^2\left(\gamma_i(\omega)-\gamma_i(-\omega)\right)=0$ for $\gamma_i(\omega)\rightarrow\gamma_i^c(\omega)$. In this sense, the quantum bath recovers the classical description in the high temperatures limit, when the spontaneous decay becomes negligible as compared to thermal effects, $n^{(i)}_\omega+1 \approx n^{(i)}_\omega$. 

The steady state driven by the classical reservoirs, which can be calculated by Eq. (\ref{rhoss1}) using the classical transition rates, is  $\rho_{ss}^c=\frac{1}{2}\left(\ket{1}\bra{1}+\ket{2}\bra{2}\right)$. Since $\rho_{12}=0$, the energy current associated with the classical dephasing,
\beq
J_1^c=
-\frac{\gamma^c_1(\omega)}{\alpha_+ \alpha_-}\left(\epsilon_+-\epsilon_-\right)\rho_{12},
\label{cheatcurrent}
\eeq
vanishes in the stationary regime. In other words, the quantum nature of the reservoir is essential to the existence of stationary energy current. It is important to mention that as the classical reservoir is not necessarily a thermal reservoir, the energy current $J_1^c$ does not necessarily describe a heat current.

%
%
\section{Conclusions}
In summary, we have shown the existence of quantum transport of heat in the steady-state regime, between two pure-dephasing reservoirs, each coupled locally to a single site. An effective system-bath energy-exchange Hamiltonian has been derived. A microscopic modeling of a quantum master equation, valid in the ultrastrong inter-site coupling regime, has been applied, yielding an effective decay rate for the chain. An effective spectral density has been identified, in analogy to the so-called Purcell effect. The transient regime has evidenced the dynamical onset of quantum coherence induced by the baths. 
Steady-state heat current has been obtained as a product between the inter-site quantum coherence and the gradient of quantum average bath excitations. The plots evidence that heat current saturates for arbitrarily high temperature gradient and has a maximum value with increasing inter-site coupling.
In the case of equal temperatures, heat current vanishes and the chain gets in a thermal equilibrium state.
Finally, it has been shown that the local approach is only valid at high temperatures and that a classical bath does not provide coherence, so heat current vanishes for classical pure-dephasing reservoirs.

An interesting perspective offered by this work is to investigate how the different types of inter-site connection in a bigger chain affect energy flow. That could be applied to microscopically model photosynthesis \cite{nori,NJP.castro,NJP.Plenio}, where the unidirectional excitation flow is still not yet fully understood.
Further consequences of the analogy to the Purcell effect could also be explored, by modeling other types of system-reservoir coupling, for instance.

%
\begin{acknowledgements}
We gratefully thank Marcelo Marchiori and Marcio Cornelio for insightful discussions.
TW and DV acknowledge financial support from CNPq, Brazil.
\end{acknowledgements}

%
%

\end{document}